\documentclass[sigconf]{acmart} 

\AtBeginDocument{%
  }

\copyrightyear{2024}
\acmYear{2024}
\acmDOI{XXXXXXX.XXXXXXX}

\acmISBN{xxx-x-xxxx-XXXX-X/xx/xx}
\acmConference[JCDL '24]{ACM/IEEE Joint Conference on Digital Libraries}{December 16--20,
  2024}{Hong Kong}

\begin{document}

\title{An Overview of zbMATH Open Digital Library}

\author{Madhurima Deb \orcid{0000-0001-5988-5366}, Isabel Beckenbach \orcid{0000-0001-6691-7362}, Matteo Petrera \orcid{1234-5678-9012}, Dariush Ehsani \orcid{0000-0001-8742-5156}, Marcel Fuhrmann \orcid{1234-5678-9012}, Yun Hao \orcid{0000-0003-2508-2017}, Olaf Teschke \orcid{0009-0003-4089-9647}, Moritz Schubotz \orcid{0000-0001-7141-4997}}

\affiliation{%
  \institution{FIZ Karlsruhe Leibniz Institute for Information Infrastructure}
  \city{Berlin}
  \country{Germany}
}
\email{madhurima.deb@fiz-karlsruhe.de}
 
\renewcommand{\shortauthors}{Deb et al.}

\begin{abstract}
  Mathematical research thrives on the effective dissemination and discovery of knowledge.
 zbMATH Open has emerged as a pivotal platform in this landscape, offering a comprehensive repository of mathematical literature. Beyond indexing and abstracting, it serves as a unified quality-assured infrastructure for finding, evaluating, and connecting mathematical information that advances mathematical research as well as interdisciplinary exploration. zbMATH Open enables scientific quality control by post-publication reviews and promotes connections between researchers, institutions, and research outputs. This paper represents the functionalities of the most significant features of this open-access service, highlighting its role in shaping the future of mathematical information retrieval.
\end{abstract}



\maketitle

\section{Introduction}
Otto Neugebauer and Richard Courant established \textit{Zentralblatt für Mathematik und ihre Grenzgebiete} in 1931. This platform was designed to provide researchers with comprehensive bibliographic information and scholarly reviews of contemporary mathematical publications from around the globe \cite{gobel2011glimpses} \cite{ehsani2020road} which aligns with the ultimate goal of making mathematical knowledge including mathematics literature, mathematical research data, and software more findable. The \textit{Zentralblatt} database became accessible via the internet in 1996 and simultaneously it was known as MATH and then ZBMATH or Zentralblatt MATH. The transition from zbMATH to zbMATH Open took place in 2021 while achieving the open-access status. 
Over the last 90 years, the technology and organization of zbMATH have changed, but the overall mission has remained constant. Today zbMATH Open covers bibliographic data and content details of publications across all branches of mathematics and its applications.

zbMATH Open offers one of the most diligent databases for bibliographic metadata in mathematical scholarly literature. It is a comprehensive provider of mathematical publications, author information, references, and mathematical software. Earlier, the usage of this wide range of information was a constraint of licensing, but since 2021, zbMATH has become open access. A high degree of formalization for interlinking the massive amount of mathematical research data has been taken up by zbMATH Open \cite{hulek2020transition}. Presently, it has indexed approximately 4.9 million publications, where the publication year spans between 1755 to 2024. zbMATH Open has been experiencing an annual growth of addition of around 120,000 new records \cite{mihaljevic2014author}. The goal of zbMATH Open is to provide all peer-reviewed mathematical research data. Under this category, many domains provide mathematical research data and other relevant resources, such as mathematical software, formulae, reviews, serials, and mathematical item classifications. It is a great source of sustainable information to the public \cite{wu2020scholarly}. Key advancements at zbMATH Open include the incorporation of nearly 800,000 full-text mathematics articles from arXiv, the expansion of author profiles, the development of a framework to disambiguate institutions, and the integration of external research data from OEIS, DLMF, and the swMATH software database.

Significant enhancement in digital library accessibility through socialization, and integration can promote literature-data-driven science \cite{hull2008defrosting}. Research on digital libraries has always been propelled by innovation, development, and a societal imperative \cite{aboelmaged2023scientometric}. 
 A mathematical knowledge corpus grows on community-based editorial contributions combined with machine learning methods \cite{national2014developing}. 
Keeping these things in mind, zbMATH Open has further networked with MaRDI\footnote{\url{https://www.mardi4nfdi.de/about/mission}} ((Mathematical Research Data Initiative), an NFDI (National Research Data Infrastructure) consortium to contribute to the development of an all-in-one platform for mathematical research data and services. zbMATH Open has strong connections with the FAIRCORE4EOSC\footnote{\url{https://faircore4eosc.eu/}} project as well, that develops and implements core infrastructure components for the European Open Science Cloud (EOSC).

\begin{figure*}
\centering
  \includegraphics[width=\linewidth]{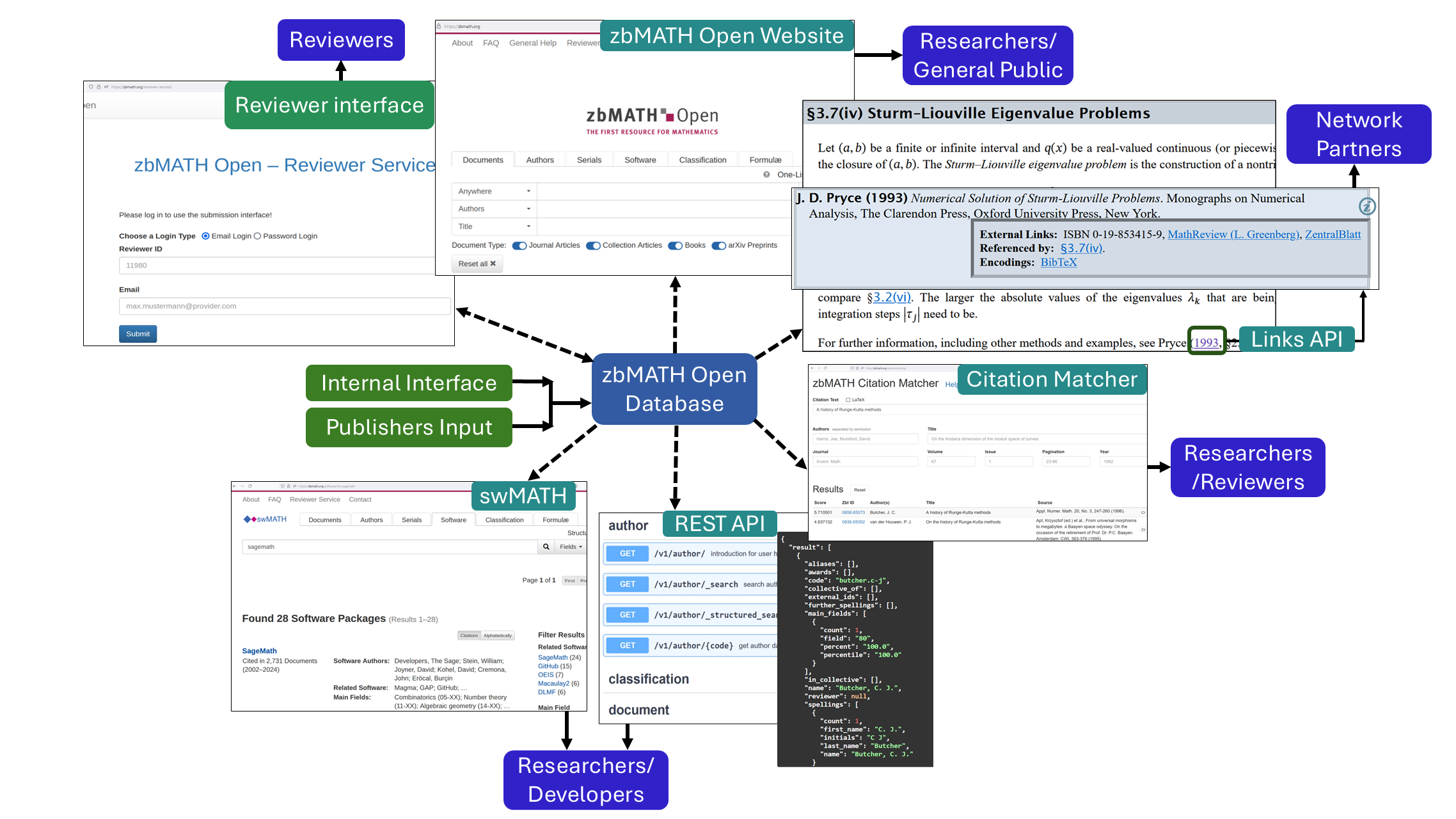}
  \caption{Visual representation of the features of zbMATH Open discussed in detail in this paper}
  \Description{This is a figure that gives an overview of the components of zbMATH Open that we are going to highlight in this paper}
  \label{fig:1}
\end{figure*}

Transforming zbMATH Open into a globally connected digital library requires offering outstanding data and services to the mathematics community. In this overview paper, we showcase some of the most recent advancements in that path.

zbMATH Open supports the trend of supplementing traditional publications with software, datasets, and relevant content for community platforms. This information is indexed based on their scientific criteria and they are available via the highly networked services provided by zbMATH Open. The interconnected services make our platform stand out as a mathematical information provider that benefits consumers immensely. Additionally, the review service is a crucial medium to offer community engagement.

We embark on a journey to unravel the rich tapestry of zbMATH Open, exploring its multifaceted functionalities such as citation matching, implementation of Links API and REST API, and their profound implications for the mathematical research community. The objective of this overview paper is to represent the technical aspects behind the zbMATH Open components that make this open-access ensemble of bibliographic metadata lucrative for the research community. The metrics for the study of the usage of zbMATH Open and its OAI-PMH API are presented in this paper. Such studies are key points for bibliometric analysis of any digital information provider \cite{hulek2024bibliographical}. We aim to illuminate not only the present landscape but also the promising horizons it opens for the future of mathematical inquiry and collaboration.

The paper is structured as follows: the related work is presented in section 2, followed by the citation matching interface at zbMATH Open named Marebito in section 3, multiple APIs at zbMATH Open including OAI-PMH API, Links API, and the REST API in section 4. Section 5 covers some additional features: the software segment, swMATH, the institute disambiguation framework, and quality control at zbMATH Open. In section 6, we present a brief analysis of zbMATH Open consumers over the past few years and shed some light on the worldwide distribution of zbMATH Open users before and after the achievement of the open-access status. It is followed by the sections covering discussion on the presented features, conclusion, and outline for future endeavours at zbMATH Open.

\section{Related Work}
The interconnected network of mathematical publications, indexing of carefully checked materials, well organized disambiguated catalog are some key characteristics of any abstracting and indexing resource for related mathematical information \cite{bouche2010digital}. This paper showcases the most significant developments at zbMATH Open, intended to support the work of individual researchers, research institutions, librarians, information professionals, and educators.

 A contemporary digital library CiteSeerX, which specializes in literature on computer science, has reported on their recent advances in functionalities such as information extraction (IE), document deduplication, author name disambiguation and key phrase extraction in \cite{wu2019citeseerx}.  Another widely accepted source for mathematical bibliographic metadata is MathSciNet\footnote{\url{https://mathscinet.ams.org/mathscinet/publications-search}} offered by the American Mathematical Society (AMS). The services provided by MathSciNet are chargeable, which is not the case for zbMATH Open. The European Digital Mathematics Library (EuDML) is another contemporary to zbMATH Open, which is more focused on European full-text collections \cite{stanchev2015presenting} and also interlinked with zbMATH Open \cite{Rakosnik2012EuDML}. 

The particular aspect that strengthens zbMATH Open is its collection of historical publications. Due to the inclusion of the Jahrbuch \"uber die Fortschritte der Mathematik \cite{teschke2021JFM}, that can be dated back to 1868. zbMATH has co-existed with the evolution of \textit{Mathematical Reviews} \cite{gobel2011glimpses} since 1940; notably, both services maintain the Mathematics Subject Classification \cite{fraser2020mathematics} \cite{schubotz2020automsc}. The literature provided by the parent of zbMATH, Zentralblatt has always been useful for tracing new work and evaluating the literature and it became more prominent starting from the initial years of digitization. In recent years, zbMATH has been focused on the objective of making scientific data (publications, software, bibliometric information) FAIR (findable, accessible, interoperable, and reusable). Open access to zbMATH substantiates connection to zbMATH Open data from anywhere around the world at any time. One of the essential characteristics of a successful digital library is determined by its rapid dissemination of knowledge and zbMATH Open demonstrates this by offering open access to its features and functionalities. This is one of the foremost reasons for the expansion of the user base of zbMATH Open. Owing to the free access, countries like the USA, Canada, and Japan have seen significant increases (almost double) in the number of users. Historical coverage, comprehensive reviews, and open access functionalities come together and make zbMATH Open the potential best choice for the community.

The utility of a digital library is heightened when the information available on that platform is findable and interoperable. Back in 2015, it was reported that zbMATH offers structured search to its users and the consumers can access 1,650,000 direct links to electronic versions of the indexed publications \cite{stanchev2015presenting}, a figure which exceeds now more than 4 million \footnote{https://zbmath.org/about/}. Users generally get access to the records via the zbMATH Open website. At the same time, making the information available in machine-readable format is equally important for the integration of the data for cross-platform usage, text mining, analysis of metadata, and even for the development of other APIs for the digital library.\footnote{\url{https://digitalcommons.elsevier.com/en_US/digital-commons-api-getting-started}} zbMATH Open has considered these points and developed two APIs that resulted in improvement of the visibility and, thus, findability of publications. As mentioned in \cite{tomaszewski2021study}, databases are nothing but pathways for discovering scientific information in diverse disciplines. The digital library REST API gives access to the data in machine-readable format. Such resource-centric RESTful services play a valuable role in integrating diverse information resources within a shared virtual infrastructure \cite{powell2005service}. REST principles were applied to establish persistent identifiers at the Cooper Hewitt National Design Museum, facilitating the reconciliation of their metadata with other knowledge bases \cite{verborgh2015fallacy}. Interestingly, a REST API named ‘RO API’ served the purpose of RO (Research Object) storage and retrieval, establishing the formats and links for the research objects, in the digital library RODL. Similarly, the ‘RO Evolution API’ was developed for systematic indexing and retrieval of workflow-centric research objects in the digital library RODL \cite{palma2013digital}. The REST API from Elsevier, known as Digital Commons API\footnote{\url{https://www.elsevier.com/products/digital-commons}} uses HTTP commands to retrieve the content. The development of RESTful API for an integrated platform like zbMATH Open contributes as a stepping stone for providing the highly linked bibliographic metadata without disturbing the existing architecture.\footnote{\url{https://digital.library.unt.edu/ark:/67531/metadc406332/}}

Bibliographic reference matching is one of the most important components of any bibliographic database and the basis of many further functionalities. Web of Science \cite{stahlschmidt2022indexation}, Google Scholar, and CiteSeerX are some of the most influential contributors in the domain of bibliographic metadata curation, along with open citation providers such as Crossref, OpenCitations\footnote{\url{https://opencitations.net/}}, and Wikicite\footnote{\url{https://http://wikicite.org/}}. An example of early work on citation matching/citation parsing is the automatic citation parsing system ParCit for citation and citation context extraction \cite{wu2019citeseerx} \cite{williams2014scholarly}, introduced by CiteSeer in 1998 which could identify citations of variable format as well as the citation context in the domain of computer science \cite{giles1998citeseer}. The citation matcher at zbMATH Open has been developed for quite some time and it has been improved over the last few years. We discuss its functionality briefly in this overview paper.

Our objective of discussing in detail the components and diverse API solutions offered at zbMATH Open is to empower the librarians and researchers with a comprehensive understanding of the platform, ultimately making it easier for them to utilize its full potential. zbMATH Open in an ensemble of inputs from publishers, reviewers, and its internal interface. The zbMATH Open website is publicly available to the global audience. Researchers and developers can easily access the swMATH interface and REST API. The citation matching interface is of particular interest to the researchers and reviewers, as is the Links API to the network partners. The reviewer interface connects the vast reviewer community to this platform. Figure \ref{fig:1} delivers the visual representation of the various interfaces of zbMATH Open highlighted in this overview paper.

\section{Marebito: Citation Matching Interface in zbMATH Open}

Researchers, funding agencies, and institutions rely on bibliographic data like citations for the evaluation of the entity, be it the impact of the research, the reputation of the research institute, or the publication venue \cite{shakeel2022altmetrics}. Citation extraction and citation matching are critical components of many research pipelines.
 By accurately identifying and linking citation information, researchers can effectively address challenges such as author disambiguation, self-citation analysis, and the creation of high-quality datasets for digital libraries. Generation and interpretation of bibliometric data is a challenging job and it is of utmost importance to highlight the technical features of the citation matching interface at zbMATH Open \cite{hulek2024zbmath}.
\begin{figure*}
\centering
\includegraphics[width=\linewidth]{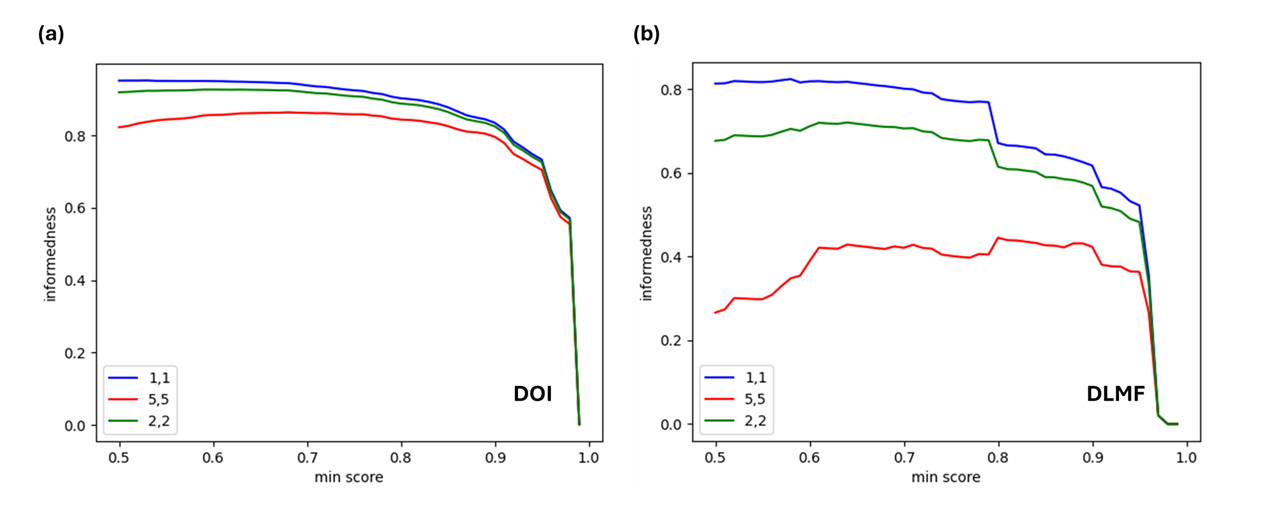}
\caption{Informedness for the citation matcher on DOI dataset and DLMF dataset}
\Description{This figure presents the informedness vs. Minimum score graph over two datasets DOI and DLMF}
\label{fig2}
\end{figure*}
zbMATH Open offers 50 million citations for nearly 2 million documents that make zbMATH Open largest citation database that has been curated for mathematics \cite{hulek2024zbmath}. The architecture of zbMATH Open’s citation matching interface internally named "Marebito" is presented in this section. A user can provide citation strings and the citation matching algorithm attempts to match the reference citation strings to zbMATH Open bibliographic entries. A reference citation string or some structured information of the publication is required as input to the citation matcher. The zbMATH Open entry is shown as output.

The machine learning library called Grobid\footnote{\url{https://github.com/kermitt2/grobid}} is used for the extraction of citation information (title, journal name, author name, year) from the input strings. The matching algorithm in zbMATH Open creates a set of candidate records for each input item. The text-based search engine Elasticsearch is used to generate the candidate items using the title and the author names of the input items. Further, similarity measuring features between each of the candidates and the input item are generated. A binary classifier aggregates these features into a single score. When the score is above a certain threshold, the matching result is accepted.

For the gold dataset, we have N test pairs. The entire test dataset is split into real positive (RP) and real negative (RN) elements. The zbMATH Open identifier used for this method is the DE number. The entries of the test dataset having a valid DE number fall under the real positive category and the entries with no DE number fall under the real negative category. Real positives (RP) are the combination of the following three categories: True Positive (TP) (entries where the DE number returned by the citation matching algorithm matches the correct DE number of the item), False Match (FM): entries that are matched to an incorrect zbMATH identifier (DE number) and False Negative (FN) (no match was found). The gold dataset generated is divided into a training set, an evaluation set and a test set. The classifiers are trained on the training set, and compared using the evaluation set. In our citation matching tool, the classifier is a trained binary classification model. 

After evaluating three classifiers on the evaluation set: SGD (Stochastic Gradient Descent), Decision Tree and Random Forest from Sklearn,\footnote{\url{https://scikit-learn.org/stable/api/index.html}} currently Random Forest classifier is used as it exhibited the best performance.

\subsection{Informedness: Performance Metric for the Citation Matching Interface Marebito}
Informedness is an evaluation measurement to quantify or evaluate the effectiveness of a classification model,\footnote{\url{https://rvprasad.medium.com/informedness-and-markedness-20e3f54d63bc}} that determines how informed the system is about the positive cases and the negative cases by calculating the balance between the tpr (true positive rate) and fpr (false positive rate). \textit{informedness} is used as the performance metric for the citation matcher at zbMATH Open.  After calculating other metrics such as precision, recall, and ROC for the citation matcher it was decided to adhere to \textit{informedness}. In the evaluation dataset, there exist varying rates of real positives (RP) and real negatives (RN) \cite{muller2016progress}. For such a skewed dataset, \textit{informedness} is an ideal and unbiased metric.

In the zbMATH Open citation matching interface, Marebito, the formula for \textit{informedness} is modified slightly from \cite{muller2016progress} to take into account all the relative penalties that are assigned to the false negatives, false matches, and false positives. The formula defining \textit{informedness}, denoted as $ \text{inf}_{\alpha,\beta}$ is given below.

\begin{equation}
 \text{inf}_{\alpha,\beta} = \frac{\mathrm{TP}}{\mathrm{RP}} - (\alpha -1)\frac{\mathrm{FM}}{\mathrm{RP}} - \beta  \frac{\mathrm{FP}}{\mathrm{RN}}  
 \label{eq:1}
\end{equation}


\equationautorefname{1}, can be written as, \begin{equation}
 \text{inf}_{\alpha,\beta} = 1- \frac{\mathrm{FN}}{\mathrm{RP}} - \alpha\frac{\mathrm{FM}}{\mathrm{RP}} -\beta\frac{\mathrm{FP}}{\mathrm{RN}}   
\end{equation}

$\alpha$ is the "cost" or "penalty" of a $\mathrm{FM}$, in the form of weight assigned to it.
 In the same way, $\beta$ is the "cost" or "penalty" of a $\mathrm{FP}$, and 1.0 is the cost of a $\mathrm{FN}$.
For zbMATH Open, false matches ($\mathrm{FM}$) or false positives ($\mathrm{FP}$) are worse than false negatives ($\mathrm{FN}$ = missing matches). As a user usually recognizes a false match more than a missing match, more penalty is assigned to $\mathrm{FM}$ as compared to $\mathrm{FN}$.
The \textit{informedness} of the citation matching tool, Marebito is calculated over two datasets. The zbMATH Open dataset, denoted as 'DOI' contains 198355 items and this dataset consists of citation matching results on zbMATH Open articles. The DLMF dataset, denoted as 'DLMF' contains 2963 items. DLMF is the dataset that contains citation matching results on DLMF articles. $\text{inf}_{1,1}$, $\text{inf}_{2,2}$ and $\text{inf}_{5,5}$ are denoted as the \textit{informedness} with both $\alpha$ and $\beta$ set to 1, 2 and 5 respectively. 
We have plotted \textit{informedness} vs. minimum score in the range of 0.5 to 1, for the above-mentioned two datasets, as shown in Figure~\ref{fig2}. We observed that for both datasets, tuning the values of $\alpha$ and $\beta$ can significantly alter the \textit{informedness}, as they are the weighting factor for the penalty. As the penalty was increased, the informedness value decreased. The \textit{informedness} for the DOI dataset was found to be above 0.8 for nearly the entire range. This indicated that the classifier was evaluated to be very efficient. For the DLMF dataset, the \textit{informedness} was lower.

\section{zbMATH Open APIs: Gateway To Bibliographic Metadata}

Several APIs ((Application Programming Interface)) in digital libraries can lead to enhancement in their functionality and accessibility, thus improving data interoperability \cite{tzouganatou2021complexity}. It surely enhances data reusability for researchers and developers and also facilitates the integration of other digital platforms and services 
within the same ecosystem. For example, CiteSeerX adopted an architecture that combines the Information Extraction and Ingestion module and uses Elasticsearch for ingesting new data. CiteSeer was one of the pioneers in developing an API for providing access to its Document, Citation, and (Citations) Group in the XML schema encoding structure. This facilitated services such as document full-text search, citation full-text search as well as retrieving the most recent collection of documents \cite{petinot2004service}. In this section, we discuss the functionalities of the various APIs developed at zbMATH Open, which are considered to be the gateway to the vast bibliographic metadata offered by this platform.
\subsection{OAI-PMH API for Bibliographic Metadata Representation}

The first goal of the implementation of the API was to offer machine-readable tools to the community so that the community can take advantage of open access to the zbMATH Open data more effectively. The second goal was to record the dynamic interactions taking place between zbMATH Open data and the data from other digital resources that are being used by the community. A standardized machine-readable format is essential for the annotation of the links between different data sets in a collection as huge as zbMATH Open. The maintenance of a Scholix-compliant\footnote{\url{https://scholexplorer.openaire.eu/}} framework is ensured for exporting data and the links between various datasets \cite{cohl2021connecting}. 

Standardized protocols enhance intradomain information system functionality and interoperability by simplifying metadata management \cite{dutta2022towards}. The system at zbMATH Open utilizes the protocol OAI-PMH, i.e., Open Archives Initiative Protocol for Metadata Harvesting, to provide synergistic assistance to potential users. OAI-PMH has been used in several projects to extract metadata in domains like cultural heritage \cite{siqueira2022workflow}, atmospheric sciences \cite{devarakonda2011data} and biomedical repositories \cite{singh2005building}. The standardized functionality of this protocol is easily accessed by motivated users. The zbMATH OAI-PMH API has six endpoints. The standard Dublin Core format is most prevalently supported for representation of the metadata. Apart from that, an additional customized format is used, especially to incorporate the MSC code \cite{arndt202110} to ensure that all the legally allowed data is covered by this XML-based API. Sometimes, owing to licensing constraints, some third-party information such as abstracts are not available to the public. The reason for opting for an OAI-PMH API is that the user needs minimal implementation efforts for accessing/harvesting the metadata from the remote service \cite{fuhrmann2023rest}. zbMATH Open started to support this protocol in the year 2020. The information systems containing bibliographic data to scientific publications are known as bibliographic consumers and they are one of the most prominent users of zbMATH Open API \cite{schubotz2021zbmath}. 

\begin{figure}
\centering
  \includegraphics[width=\linewidth]{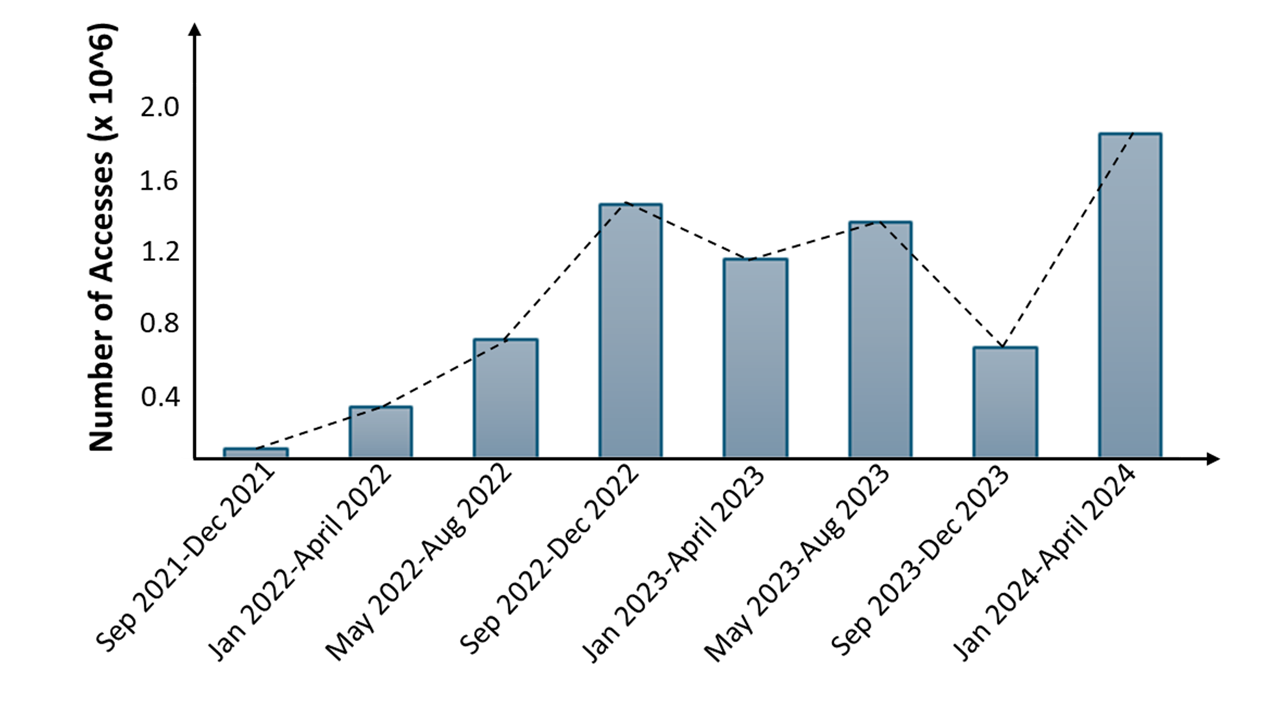}
  \caption{Study of estimated yearly accesses of zbMATH Open OAI-PMH API for the three years 2021, 2022 and 2023}
  \Description{The comparative study shows a bar chart of yearly number of accesses in the digital library zbMATH Open for three years 2021, 2022, 2023}
  \label{figure 3}   
\end{figure}

  We have taken the opportunity to represent the number of accesses for the zbMATH Open OAI-PMH API starting from September 2021 to April 2024. The data has been retrieved from.\footnote{\url{https://oai.zbmath.org/cgi-bin/awstats.pl}}\textsuperscript{,}\footnote{\url{https://zbmath.org/cgi-bin/awstats.pl?month=all&year=2021&output=main&config=oai.zbmath.org&framename=index}} The number of accesses has seen almost a five-fold increase from September 2021 to April 2024 as shown in Figure \ref{figure 3}. Especially, the growth in the usage of the OAI-PMH has been monotonic from 2021 to 2022. 

\subsection{Links API}

Links API is the connection between the database of zbMATH Open and external platforms displaying the mathematical open-access research articles indexed at zbMATH Open. The Links API started with hosting one partner DLMF.\footnote{\url{https://dlmf.nist.gov/bib/}} The DLMF bibliography scraping resulted in the dataset for this Links API. At this time, the API hosted a total of 2053 zbMATH Open-indexed references and 6526 links. In this context, DLMF, the external partner is described as the source, and zbMATH Open is described as the target. Thus, the links from DLMF are termed as source objects and the zbMATH objects are target objects. The endpoints of the Links API are listed in \cite{petrera2024api}.
 This API is beneficial to a plethora of users such as digital archives, semantic data extractors, internet search engines, and bibliographic information providers. Data integration of DLMF has been successfully done into the zbMATH Open database, which is tremendously useful to mathematics scholars. For each citation instance of every article in DLMF, an individual link is created in zbMATH Open. The JSON output is structured on the Scholix metadata schema. The Links API or the Scholix Link API provides the interconnection between zbMATH Open and its partner DLMF. The 36 chapters and 75\% of its 2800 bibliographic references are directly linked to zbMATH Open \cite{ehsani2023integration}.

An important utility of Links API is that it gives the user a platform to get a coherent understanding of the development of a scientific topic by tracing the bibliographic content over some time. The Links API has made it possible to form a connection between the MSC code and the DLMF Links that enables more advanced bibliographic search by the users. 
Recently, another source called the Online Encyclopedia of Integer Sequences (OEIS)\footnote{\url{https://oeis.org/}} has been interlinked with the zbMATH Open database. OEIS is an interactive mathematical service covering calculations of data, that contains both data and tools \cite{sperber2016mathematical}. This platform holds an online database of sequences of numbers. This collection contains a total of 342,422 sequences and each of these sequences is associated with their own set of metadata such as initial elements of the sequence, formulas for generating the sequence, citations to books, articles, and scholarly references where the sequences have been featured, along with additional details.

Another new partner in the domain of group theory, ATLAS\footnote{\url{https://brauer.maths.qmul.ac.uk/Atlas/v3/}} is in the process of getting connected with zbMATH Open. Nearly 300 new links from 26 sporadic groups have been added already that can be accessed from the swagger interface of the Links API.

 The platform for mathematics question-answer from the Stack Exchange network, MathOverflow\footnote{\url{https://mathoverflow.net/}}, which is already connected to zbMATH Open is a future partner for the Links API. This is a big step towards improving the interoperability and findability of mathematics research. In a bidirectional linking framework, on the one hand, zbMATH Open records are linked to MathOverflow questions using their Stack Exchange API and on the other hand, zbMATH Open indexed items can be directly cited on MathOverflow \cite{beckenbach2024zbmath}. The search results on the MathOverflow side also include preview links to the articles at
zbMATH \cite{muller2019references}. Mathematics education, which is getting more and more dependent on online resources \cite{hasumi2022online}, will benefit from this partnership between MathOverflow and zbMATH Open. MathOverflow has gained a lot of answers using the zbMATH Open author database and several cases of academic misconduct and plagiarism have been identified from the MathOverflow platform \cite{muller2019references}.

The Links API has a multitude of diverse use cases. It is widely usable in tracing bibliographic references related to a certain topic under the present partners of zbMATH Open, DLMF, and OEIS. This API enables a researcher to verify for a zbMATH Open object cited at DLMF or OEIS, furthermore, one can retrieve the links associated with a mathematical subject classification (MSC) code or a particular author ID. 6312 links from zbMATH Open to DLMF and 67,436 links from zbMATH Open to OEIS are displayed on the zbMATH Open website. This is an essential tool for statistically analyzing the bibliographic content in DLMF or OEIS.
  It was found that the most cited primary MSC codes for DLMF are 33 (Special functions), 65 (Numerical Analysis), and 11 \cite{petrera2021zbmath} and the most cited primary MSC code in OEIS is 11 (Number Theory) \cite{ehsani2023integration}. Another additional statistical analysis provides insight into the most frequent publication years as per the cited references \cite{ehsani2023integration}. Apart from this, the LinksAPI has made it possible to provide the URL to the external partner (OEIS and DLMF) for a particular item indexed in zbMATH Open. 

\subsection{zbMATH Open REST API}

\begin{figure}
\centering
\includegraphics[width=\linewidth]{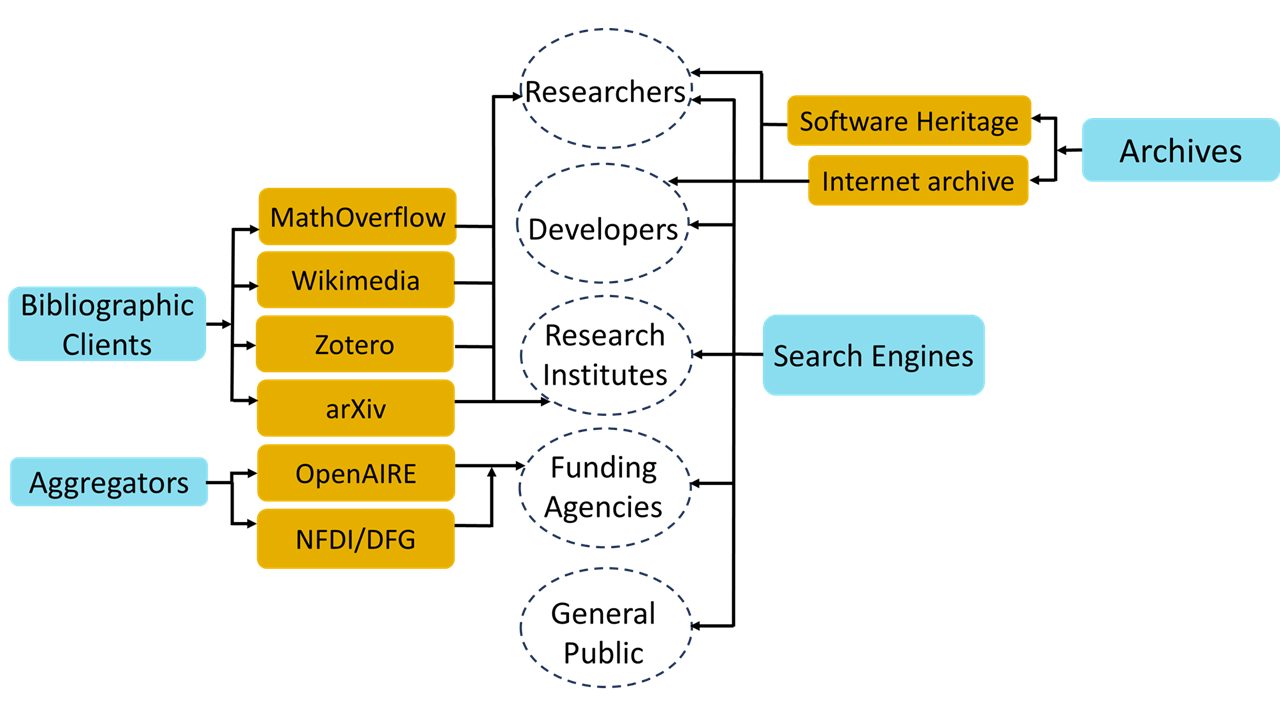}
\caption{The diverse user groups of zbMATH Open}
\Description{A wide range of users are benefited from the information provided by zbMATH Open. In this figure, we have consolidated our potential users in one place}
\label{figure 4}
\end{figure}

The disadvantages posed by OAI-PMH, such as only XML-based data availability, limited metadata in the standard format, and limited filtering criteria (time and subset only) led to the establishment of a REST API for zbMATH Open \cite{fuhrmann2023rest}. The REST API provides machine-readable content from zbMATH Open. The cacheable data is distributed under the license CC-BY-SA 4.0.\footnote{\url{https://api.zbmath.org}} Depending on the publisher agreement, the REST API is capable of automatically redacting partial information (like abstracts or references) from articles. The structured search is a convenient search option for all relevant fields whereas the free logical combination of available search fields can be done via syntax search. The syntax search offers more flexibility than the structured search. The author endpoint provides details of the authors. The classification endpoint provides information about the MSC code of a particular research field. The document endpoint can be used to retrieve metadata information regarding the specific documents in zbMATH Open following various search fields. The serial endpoint can be utilized to find information on journals or book serials. The software metadata can be retrieved from the software endpoint. It is interesting to note that the REST API was added in 2023 and it has more than 300,000 accesses in the year 2024 itself. 

zbMATH Open grants access to its mathematical research-based services to individual researchers as well as its potential consumers such as search engines (Firefox search plugin), bibliographic clients (MathOverflow, Wikimedia, arXiv, Zotero), archives (Software Heritage, Internet archive), and aggregators (OpenAIRE/European Research Council, NFDI/DFG)  \cite{schubotz2021zbmath}. We have shown in Figure \ref{figure 4} the diverse user groups of zbMATH Open. Researchers and research institutes leverage bibliographic clients, funding agencies utilize aggregators, researchers, and developers benefit from archives, and users from all these categories utilize search engines to access zbMATH Open services. 

\section{Additional Features at zbMATH Open}

Some additional components of zbMATH Open are highlighted in this section. Institutional disambiguation framework development is one of the most recent advancements of zbMATH Open. This section also puts forward a brief discussion on the software metadata retrieval platform swMATH and quality control at zbMATH Open.

\subsection{Institutional Disambiguation}

An internal framework for institute disambiguation is being developed at zbMATH Open. This framework employs both algorithmic and manual curation techniques, utilizing information from zbMATH Open and other open data sources. The process involves matching raw affiliation data against an in-house database, which uses the ROR\footnote{\url{https://ror.org}} (Research Organization Registry) database as a foundational resource. This in-house database also extends to include specific departments within institutes. The matched affiliations are further linked to additional open datasets, such as OpenAlex\footnote{\url{https://openalex.org/}} and Wikidata\footnote{\url{https://www.wikidata.org/wiki/Wikidata:Main_Page}}. Notably, the ROR database indexes approximately 109,000 institutes, with around 30\% of these institutes producing mathematics-related papers indexed by zbMATH Open. Currently, zbMATH Open contains nearly 600,000 indexed items with associated affiliation data, most of which is supplied by publishers. This is an excellent feature offering open data on disambiguated institutional entities.

\subsection{Software}

Mathematical software plays a pivotal role in the context of mathematical research data. Mathematical research data is essential for the reproduction, verification, and reuse of mathematical research results. Mathematical software, along with mathematical models and algorithms are evolving as an integral part of interdisciplinary research \cite{hulek2019mathematical}. swMATH is an information service that indexes mathematical software.\footnote{\url{https://zbmath.org/software/}}
 It is integrated into zbMATH Open and it has facilitated the discoverability of mathematical software \cite{azzouz2022sustaining}. Now zbMATH Open can offer references to nearly 40,000 mathematical software along with the associated cross-linked information such as algorithms and other functionalities. The dynamic nature of the management of mathematical research data such as mathematical software is a critical aspect of any digital information infrastructure and zbMATH Open has become quite successful in this domain by integrating the largest catalog of mathematical software, i.e., swMATH. Users can get a simple overview of the existing software packages, their possible applications, and use cases.

Evaluation of zbMATH Open bibliographic data fields such as titles, keywords, MSC codes, and citations related to mathematical software \cite{chrapary2024swmath} provides software information at swMATH. External sources such as arXiv, specialized journals, 
and dedicated websites are also explored for extracting software information. Interestingly, nearly 475,000 software references were recorded in 244,084 entries in zbMATH Open by the end of 2021  \cite{chrapary2024swmath}.
\begin{figure*}
\centering
  \includegraphics[width=\linewidth]{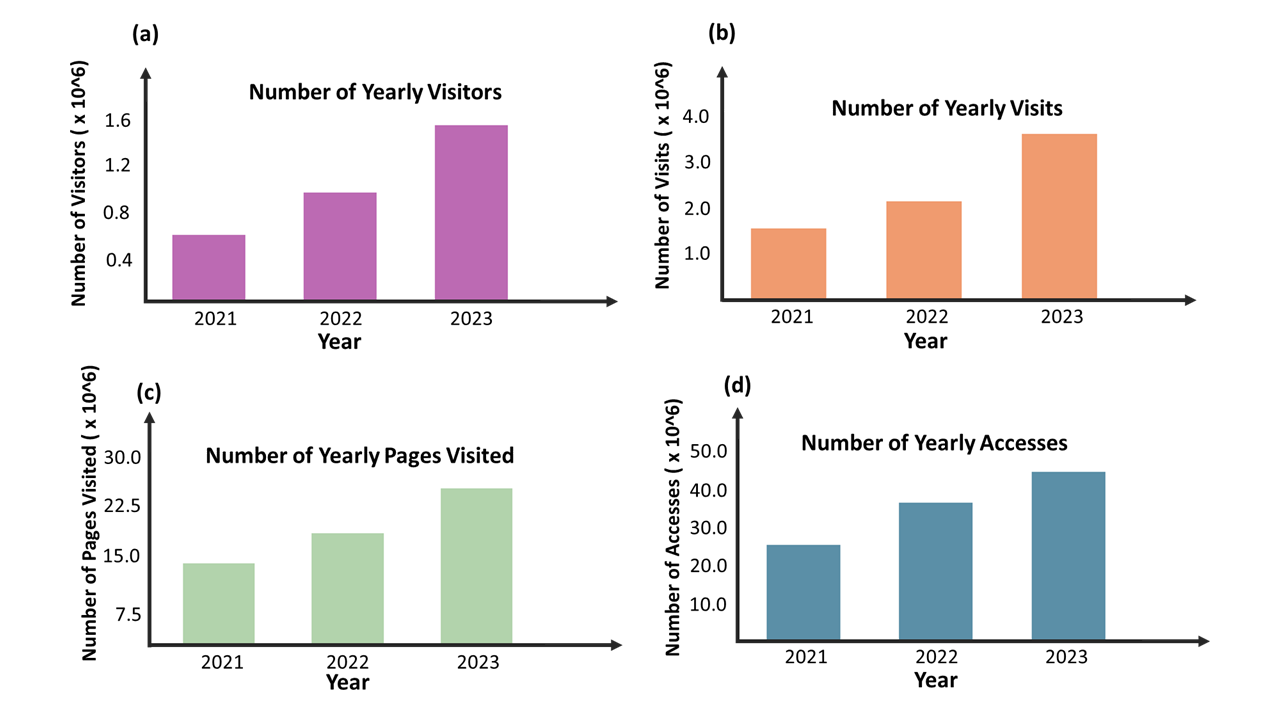}
  \caption{Comparative study of estimated yearly users of zbMATH Open for the three years 2021, 2022 and 2023}
  \Description{The comparative study shows a bar chart of yearly number of visits, yearly number of visitors, yearly number of pages visited and yearly number of accesses in the digital library zbMATH Open for three years 2021, 2022, 2023 }
  \label{figure 5}
    
\end{figure*}

 \subsection{Quality Control at zbMATH Open}

In a vast network such as zbMATH Open, maintaining quality control is of utmost importance. On the one hand, zbMATH Open fulfills the demands of a wide range of uninterrupted functions, on the other hand, it follows very rigorous selection criteria. While choosing appropriate journals to be indexed in zbMATH Open, identifying predatory journals is a crucial step. Moreover, the reviews and abstracts provided by zbMATH Open can be of significance for the consumers to identify high-quality research data \cite{werner2024quality}. 
The open availability of reviews enables wide distribution of the reviews. In addition to that, editors are appointed to monitor the reviews on this platform, making quality control a multilayered activity.

\section{Analysis of Consumer Data of zbMATH Open}
Analysis of consumer data at zbMATH Open is very crucial to understand user behaviour and preferences. It is an integral part of the process of growth for an open-access platform like zbMATH Open. This analysis paves the way towards tailored improvements and optimizations in the user interfaces. It is also essential for measuring the impact of zbMATH Open in supporting the mathematical community. This section describes the studies on yearly consumer data as well as the worldwide distribution of zbMATH Open users.

Understanding user demographics is a pivotal aspect of the growth of an information system such as zbMATH Open. We have undertaken a few studies of the consumer data of zbMATH Open for the years 2021, 2022, and 2023, as shown in Figure \ref{figure 5}. The following analysis is done on the statistics for zbMATH.org\footnote{\url{https://zbmath.org/cgi-bin/awstats.pl?month=all&year=2021&output=main&config=zbmath.org&framename=index}}.  
This demographic study is an initial step toward understanding and analyzing user behaviour. To understand how consumers interact with zbMATH Open, we have taken into account the yearly total number of visitors, the total number of visits, the total number of pages visited and the total number of accesses respectively. This study was quite useful for the visualization and comparison of consumer data over the last three years. It is interesting to note that the number of distinct visitors has seen a steep increase in the last three years, as shown in Figure \ref{figure 5}.a. The number of accesses has crossed 48 million in 2023 as shown in Figure \ref{figure 5}.d. 

\begin{figure*}
\centering
  \includegraphics[width=\linewidth]{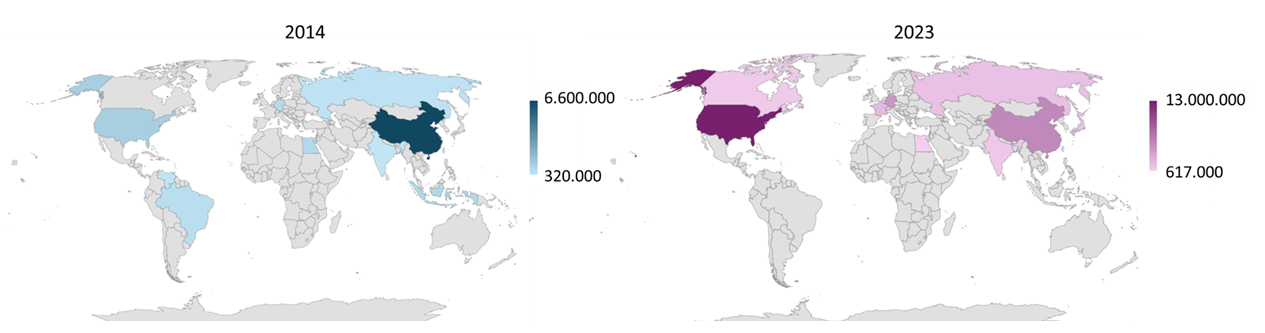}
  \caption{World wide distribution of users covering estimated number of access of zbMATH Open in the year 2014 (before open access status) and in the year 2023 (after open access status)}
  \Description{The distribution of users of zbMATH Open from all over the world in the years 2014 and 2023, based on the statistical data on the number of accesses of zbmath.org from the top ten countries}
  \label{figure 6}
\end{figure*}

We analyzed the total number of zbmath.org accesses from the top ten traffic-contributing countries for two years: 2014 and 2023 and compared the geographical distribution of users. We chose one year (2014) from the pre-open access era and one year (2023) after zbMATH was granted the open access status.
 When we look at Figure \ref{figure 6}, the highest traffic in the year 2014 came from China, followed by the USA and then Russia. In the year 2023, we observe the USA shifting to the first position and China taking the second place, followed by Russia and India. This is a very interesting study in terms of evaluating the success of the open-access status of zbMATH. From 2021 onwards, when zbMATH transitioned into zbMATH Open \cite{hulek202490,hulek2022transition}, it became available for a much larger user base all across the world. There was a lot more interest in this open-source of mathematical research data evolving in countries like Canada and Japan. We can identify this from the above study. It is crucial to point out here that a significant portion of this traffic at zbMATH Open comes from the non-subscribers owing to the greater number of searches for a much-refined result. This creates a bias in the country-wise distribution towards some countries, such as Indonesia or Egypt.

\section{Discussion}

In 2020, zbMATH achieved renewed indexing contracts that enabled Open Data services via the zbMATH Open platform that includes reviews, author disambiguation data, and their interlinked data under the CC BY-SA license.\footnote{\url{https://creativecommons.org/licenses/by-sa/4.0/}} A lot of effort is being undertaken at zbMATH Open to expand the size of this information base. The additional information layer generated by interlinking OEIS and DLMF to zbMATH Open has provided the opportunity for unconventional navigation and analysis \cite{ehsani2023integration}. The open access format enabled the integration of principles of data minimization and data avoidance \cite{hulek2020transition}.

Notably, the MSC tags and keywords for the research articles provided by zbMATH Open serve the purpose of classification as well as clustering. Building use case-specific search and clustering schemes are emerging as a possible result of open access to the APIs.
zbMATH Open thrives to improve as a modern and open reference tool for mathematical research data. It will continue to build connections with external mathematical resources and improve the visibility of the research data available on zbMATH Open. zbMATH Open contains now almost 100,000 new links to resources like Unpaywall and CiteSeerX \cite{hulek2020transition}.

Owing to the integration of mathematical articles from arXiv into zbMATH Open, the number of available full texts has gone up to over a million \cite{muller2016progress}. A detailed representation of community dynamics of the submissions of arXiv has been presented in \cite{muller2016progress}. Additionally, the connection of arXiv articles with author profiles has enhanced the discoverability of young researchers and has accelerated the dissemination of knowledge. Mathematics-related arXiv articles can be matched with zbMATH Open documents by using title, author, and abstract similarity as features, for more details see \cite{petrera2021zbmath} or \cite{beckenbach2024extension}. It is worth mentioning that, author disambiguation framework has played a significant role in this regard. Identifiers from and links to 15 external services are networked with the author profiles now \cite{hulek2020transition}. This has further opened up new avenues for bibliometric studies like co-author distance etc.
 
Additionally, ongoing initiatives strive to empower users with the ability to effortlessly incorporate or update data through the REST API. The seamless integration of this capability into the REST API aligns harmoniously to broaden access to zbMATH Open.


\section{Conclusion and Future Work}

The success of zbMATH Open has been possible by constant research and development. zbMATH Open offers vast data that can be utilized as test bed for numerous projects, algorithms, and models. Several Open Science projects have been undertaken by FIZ Karlsruhe based on the zbMATH Open data. The tools that are developed and maintained at zbMATH Open, enhance the discoverability of mathematics literature and this, in turn, boosts mathematics research and learning. The goal of zbMATH Open will always include faster and easier dissemination of mathematical knowledge using automatization, development of newer tools, and the interconnection of various platforms offering mathematical research data.

The progress of zbMATH Open has witnessed the evolution of its components. There is an ongoing effort to eliminate the present drawback of the citation matcher which involves the inability to identify the references not indexed in zbMATH Open. Author Collaboration Identification is a potential research question that can be addressed in future research. A comprehensive author collaboration graph has already been developed from the study of zbMATH Open author distance \cite{hulek2023mathematicians} and it is a direct outcome of the open access nature of zbMATH Open data. The construction of collaboration graphs delivers the opportunity to identify the collaboration effectiveness of researchers. About the institute disambiguation framework, several exciting features are on their way to be incorporated at zbMATH Open, for example, additional information on departments will be available along with the institute information, which is a unique step undertaken by any mathematical information database. In addition to that, the users will not only be able to retrieve the information but also edit it. The edits will undergo moderator review to ensure quality control and the identity change of institutions will also be taken into account with time. zbMATH Open API plans to facilitate the direct upload of bibliographic metadata of publications by publishers, upload of full-text by users, and upload of external identifiers like DOI or arXiv identifier to the already existing zbMATH Open articles \cite{fuhrmann2023rest}.

Retraction tracking is another essential aspect to be incorporated in zbMATH Open \cite{petrera2021zbmath}. As the Links API expands over more external partners, the required restructuring will be undertaken. An important future aspect of the API is to expand its communication with the MaRDI knowledge graph, and support protocols such as RDF, and SPARQL. 

In addition to that, the automatic recommendation of keywords and assignment of MSC codes fall under future objectives that can be beneficial to the mathematical community. Voluntary involvement from the mathematical community can help address many challenges in this domain, such as copyright limitations on full texts, diverse formats, and the absence of semantic information. The recommendation system for papers and improvement of search functionalities by the integration of LLMs (Large Language Models) are some of the ongoing works at zbMATH Open \cite{satpute2024can}. We also aim to explore the correlation between user demographics and the subjective relevance of mathematical topics by developing user models \cite{dahm2017vision}.

In conclusion, the ongoing enhancements and future directions outlined in this discussion signify the continuous evolution of zbMATH Open, with a commitment to overcoming challenges and providing valuable tools for the mathematical community. As we embark on this journey of innovation and collaboration, the horizon holds promising opportunities for enriching research, fostering partnerships, and advancing the accessibility and functionality of zbMATH Open for the global mathematical community.

\section*{Acknowledgements}
 This research is supported by Deutsche Forschungsgemeinschaft (DFG) under Grant Agreement No. \href{https://gepris.dfg.de/gepris/projekt/460135501?context=projekt&task=showDetail&id=460135501&}{460135501}, NFDI, 29/1 “MaRDI - Mathematische Forschungsdateninitiative” and the FAIRCORE4EOSC (Core Components Supporting a FAIR EOSC) project, funded by the EU’s Horizon Europe Research and Innovation Programme under Grant Agreement No. \href{https://cordis.europa.eu/project/id/101057264}{101057264}.

\bibliographystyle{ACM-Reference-Format}
\bibliography{zbmath}
\end{document}